\documentclass[referee]{cjaa}           

\usepackage{graphicx}                   
\input{epsf.sty}                        
\input{psfig.sty}                       

\begin{document}

   \title{May Gravity reveal Tsunami ?
}


   \author{D.Fargion
      \inst{1,3}\mailto{daniele.fargion@roma1.infn.it}
      }
   \offprints{D.Fargion}                   

   \institute{Physics Department, Rome University 1, La Sapienza, Rome\\
             \email{daniele.fargion@roma1.infn.it}
        \and
             INFN, Rome 1,Ple.A.Moro 2,00185,Rome,Italy\\
          }

   \date{Received~~2005 month day; accepted~~2005~~month day}

   \abstract{ The present gravitational wave detectors are reaching
  lowest metric deviation fields able to detect
    galactic and extra-galactic gravitational  waves,
   related to Supernova explosions up to Virgo cluster.
   The same gravitational wave detector are nevertheless almost able to reveal, in principle, near
   field Newtonian  gravitational perturbations due to fast huge mass displacements
   as the ones occurring during largest Earth-Quake or Tsunami
   as the  last  on  26nd December 2004 in Asiatic area. Virgo and
   Ligo detector are unfortunately recording on high frequencies (above tens
   Hz) while the signal of the Tsunami lay at much lower range (below
   $0.1$ Hz).   Nevertheless prompt gravitational near field deformation by the Tsunami
  might reach the future LISA threshold sensitivity and frequency windows $if$ such an  array is
  located nearby  $\simeq 3000$-$10000$ km distances. Unfortunately  the
  present LISA system  should be located at Lagrange point too far
  away ($1.5$ million km. far away).   We note however
  that the later continental   mass rearrangement and their gravitational
  field assessment on Earth must induce, for Richter Magnitude $9$ Tsunami,
   a different terrestrial inertia momentum and a different principal rotation
      axis. In conclusion we  remind that  gravitational geodetic
      deviation on new precise satellites (GOCE 2006),  assisted by
      GPS   network, might nevertheless  reach  in the near future the  needed threshold and
accuracy to reveal   Tsunami by their prompt tidal gravity field
deviations . An  array of such geoid detector with LISA-like
satellite on Earth orbits may offer the fastest alarm system.}

   \authorrunning{D.Fargion }            
   \titlerunning{Gravitational tidal waves to reveal Tsunami}  


   \maketitle
%
%
\section{Introduction}

The last lethal  Tsunami event corresponds to an energy release
equal   or greater than $E_{Tsu}\simeq$ $10^{27} erg$ ( Richter
Magnitude $9$ Tsunami) released in a very short time, shaping the
sea into a wide  spread Tsunami wave front.  These events are well
observed by normal sismic detector, whose   time response are
related by the surface  or sound propagation  on Earth. On the
contrary the huge energy release maybe in principle    source of
other signals  detectable at the faster velocity of light. Here I
consider a preliminary model associated with  the mass
displacement occurring during the largest Tsunami   as the last
huge one ; in order    to  make the approximation flexible    I
will calibrate the results following the energy release scale
$8.9$ Richter scale or $E_{Tsu}\simeq $ $10^{27} erg$. We foresee
that a detectable shrinking of the averaged Earth radius of nearly
$ \Delta {R_{\oplus}} \simeq 3.4 \mu$m. and a consequent faster
Earth spinning by a ratio   ${\frac{\Delta
w_{\oplus}}{w_{\oplus}}} \simeq - 1.08 \cdot 10^{-12} $
      had take place and a year duration suffered a shortening  of the order of  $\Delta {t_{year}} \simeq - 34 \mu $ s;
       if all the energy released is absorbed by the terrestrial
       rotation energy the opposite would occur,
       ${\frac{\Delta w_{\oplus}}{w_{\oplus}}} \simeq 2.59\cdot 10^{-10} $,
        with a longer year lenght:$ \Delta t_{year} \simeq  8.17 \cdot 10^{-3}$
       s; because the terrestrial gravitational energy is nearly $480$
       times its rotational one, the mutual energy exchange may in
       general   leads to a spin up or a  spin down  of the day
       lenght within $ \mp\Delta t_{day} \simeq  22.4 \cdot
       10^{-6}$s. well within detection.

\subsection{Energy output of Largest Earthquake and Tsunami}

   Let us remind  the huge power occurring during
  largest Tsunami or Earth-Quake, comparable to nearly one and a half million
  of Hiroshima nuclear explosions or $3.2 \cdot 10^{10}$ TNT. In other words it is like $10^4$ kilometer water cube
  are falling down from a km height, or an  ice asteroid of $0.1 km^3$ volume was hitting the Earth.
    If this event would be triggered by artificial or exotic means, as discussed below, for
example by   a ton anti-meteorite annihilation, a million nuclear
Hiroshima weapons explosion,  a ton mass mini-black-hole
evaporation, than its detection would be
  promptly achieved by its prompt neutrino emission and the warning alarm might  be immediately
  given. However the source of the Tsunami are related to
  geological elastic (and after all rotational and  gravitational) energy release
   and their prompt associated signal is nearly
  undetectable out of a tiny near field Newtonian gravitational
  perturbation.

\section{Gravitational Near Field by a Tsunami}

   Some geological ideas on geological effects observable by
   interferometer detector has been already put forward recently
   \cite{geo}.
  Here, independently, we   estimate the adimensional metric displacement by a
  Tsunami by the following approximation:  we assume that a sea
  water mass $m_w$ is raised by a Tsunami by a height $h$ (its exact value is irrelevant),
  assuming that the energy conversion from the Tsunami $\Delta E_{Tsu}$ to the
  gravitational energy is comparable (out of a factor $\eta$):
     $$ m_w = \frac{\Delta E_{Tsu}}{g h} \cdot \eta$$

  The consequent adimensional metric deviation $h_{\alpha \beta}$ along Cartesian axis along the
   unitary vectors due to the tidal mass displacement has a value:

$$ h_{\alpha \beta} =  \delta g_{\alpha \beta} \simeq 2 \frac{G m_w}{c^2 r} \frac{h}{r} \cdot \eta  $$
  Therefore the final absolute value (out of an angular projection
  not considered here) is:
$$ h_{\alpha \beta} = 2\cdot \frac{G \Delta E_{Tsu}}{c^2 r^2 g} \cdot \eta  = 1.3 \cdot \eta 10^{-22} ({\frac{10^9 cm}{r}})^2$$
  These metric perturbation are within present and future LISA  thresholds and frequency ($10^{-2}$-$10^{-3}$ Hz.). However the energy conversion
  $\eta $ maybe $1\%$ and the needed threshold is lost. However it
  is quite surprising that the newtonian tidal field is already in
  the range of values of present GW antennae. At a  nearby
  distance of $3000$ km the detection threshold is reduced to $
  10^{-21}$ well within detector thresholds.

\subsection{Terrestrial Gravitational shrinking and  angular
velocity speed-up}

  One of the long term consequence of the huge mass displacement
  inside the Earth is the change of its Inertial Momentum as well as of the terrestrial
  gravitational field; its readjustment  must compensate the energy dispersion of the Tsunami
  itself. Elementary consequence  are derived by angular
    momentum and energy conservation, once the elastic energy is replenished by the Earth gravitational one:

    $$ E_{g \oplus} = - \frac{G}{2} \cdot \frac{{M_{\oplus}}^2}{R_{\oplus}}$$
    $$ \frac{\Delta r}{R_{\oplus}}= \frac{\Delta
    E_{Tsu}}{E_{g \oplus}} \simeq -5.43 \cdot 10^{-13}\frac{E_{Tsu}}{10^{27} erg} $$

    The consequent average terrestrial radius contraction is:

    $$\Delta r \simeq   -3.4 \frac{E_{Tsu}}{10^{27} erg}\mu \cdot m $$

     The angular momentum conservation imply a faster spin
     frequency for the Earth:   $ \frac{\Delta w}{w} = 2 \cdot \frac{\Delta r}{r}$
     and a consequent  shortening of the year duration by a well
     detectable value:

     $$ \Delta t_{Tsu} \simeq -3.4 \cdot 10^{-5} \cdot \eta s.$$

     A value well within present time measure accuracy.

\subsection{Terrestrial Rotational Energy dissipation and  angular
velocity slow-down}

     There is also the possibility that  part of the energy
     released has been promptly absorbed by the terrestrial rotational
     energy $E_{Rot \oplus} = \frac{1}{2} I w_{Rot \oplus}^2$ by a drastic inertial mass redistribution;
      in this case the terrestrial angular velocity may be slow down
     by a larger fraction: $$\frac{\Delta
    E_{Tsu}}{E_{Rot \oplus}}\simeq 2.59 \cdot 10^{-10} \frac{E_{Tsu}}{10^{27} erg}$$ and our previous
    estimate reverse on a slow down and a day lenght increase  of opposite sign and nearly $2$
    order of magnitude larger. This extreme case is of great
    interest  and each year time accumulate increase  will be very soon
    detected:  $\Delta t_{Tsu} \simeq 8.13 \cdot 10^{-3} \cdot \eta s. $ while
  the corresponding day lenght will be increased by a time Tsunami delay:
  $$ \Delta t_{Tsu} \simeq 2.24 \cdot 10^{-5} \cdot \eta s.$$

  \subsection{Rotational-Tsunami Energy  Equipartition by Earth's Gravitational field}

  In some sense previous estimate are both to be considered  as an upper and a lower bound
  limits, because the energy balance budget (for either Tsunami and Terrestrial rotation) maybe covered by the
  dominant Gravitational one whose value is nearly $480$ times
  the terrestrial rotational energy.

  Therefore it is probable that  $ \Delta t_{Tsu} \simeq \mp 22.4 \cdot 10^{-6}\frac{E_{Tsu}}{10^{27} erg} \cdot \eta
  s$ a day, while a terrestrial average radius variation $\Delta r \simeq  -3.4 \cdot \eta \mu\cdot m
  $; in case of a more probable combined Earth Gravitational energy flow in equipartition both in
  Tsunami  and in the Rotation terrestrial  energy the estimated averaged Earth
  radius shrinking would be twice as large: $\Delta r \simeq  -6.8\cdot \eta \mu\cdot m
  \frac{E_{Tsu}}{10^{27} erg} $ and the terrestrial spin will be speed up by a few or ten microsecond a day ,
  depending on the exact mass redistribution.
   The nutation of the bending angle of the Earth spin axis (due to the Earth inertial mass change)
    might as large as the  same  factor
   $ \simeq 2.6 \cdot 2 \pi \cdot 10^{-10} \frac{E_{Tsu}}{10^{27} erg}$
   leading to a prompt terrestrial axis pole displacement as long
   as a cm. size; however being the whole Earth  nutation axis trajectory
   around the pole already a few tens a meter a year, this miss-alignment might be difficult to
   be disentangled, while the day period increase (or decrease) might be probed also by
  its steady grow and by the  accurate atomic and astronomical timing technique.

\subsection{Tidal  fields and  Geoid detections by GOCE satellite}
  In analogy with the near field estimate  the gravitation acceleration deviation is :

  $$\delta g \simeq 2 \cdot \frac{G \Delta E_{Tsu} }{g r^3} $$
  $$\delta g \simeq 3.56 \cdot10 ^{-9}{(\frac{3\cdot 10^3 km}{r})}^3 cm s^{-2}$$
  just at the edge ($3$ nGal) of present interferometer  thresholds for
  LISA like detectors.
  Let us mention the very recent proposal of GOCE , a satellite
  able to track at best its keplerian trajectory up to highest resolution
  by its GPS system. The exact geodetic track of GOCE allows to reveal a geoid
  deformation below a centimeter lenght.
   The Tsunami tides were comparable to 1.8 $cm$ size. Thefore future GOCE detector
    (a novel satellite able to better trace geoid perturbations) will be
     able to reveal largest Tsunami by their gravity signals , see ref.\cite{GOCE}.
     A very exciting novel use of near field gravity detector.

\begin{figure}
   \vspace{2mm}
   \begin{center}
   \hspace{3mm}\psfig{figure=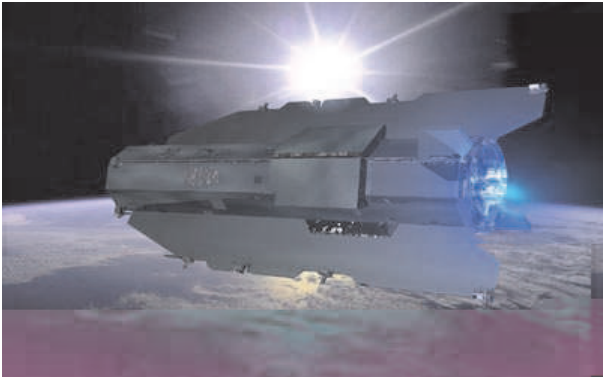,width=80mm,height=50mm,angle=0.0}
   \caption{ The on-going GOCE satellite whose wings allow (at hundred kilometer quota) to gently correct its geodetic
   for a better and more precise geoid deviations, see ref.\cite{GOCE}. }
   \label{Fig:lGOCE}
   \end{center}
\end{figure}

\begin{figure}
   \vspace{2mm}
   \begin{center}
   \hspace{3mm}\psfig{figure=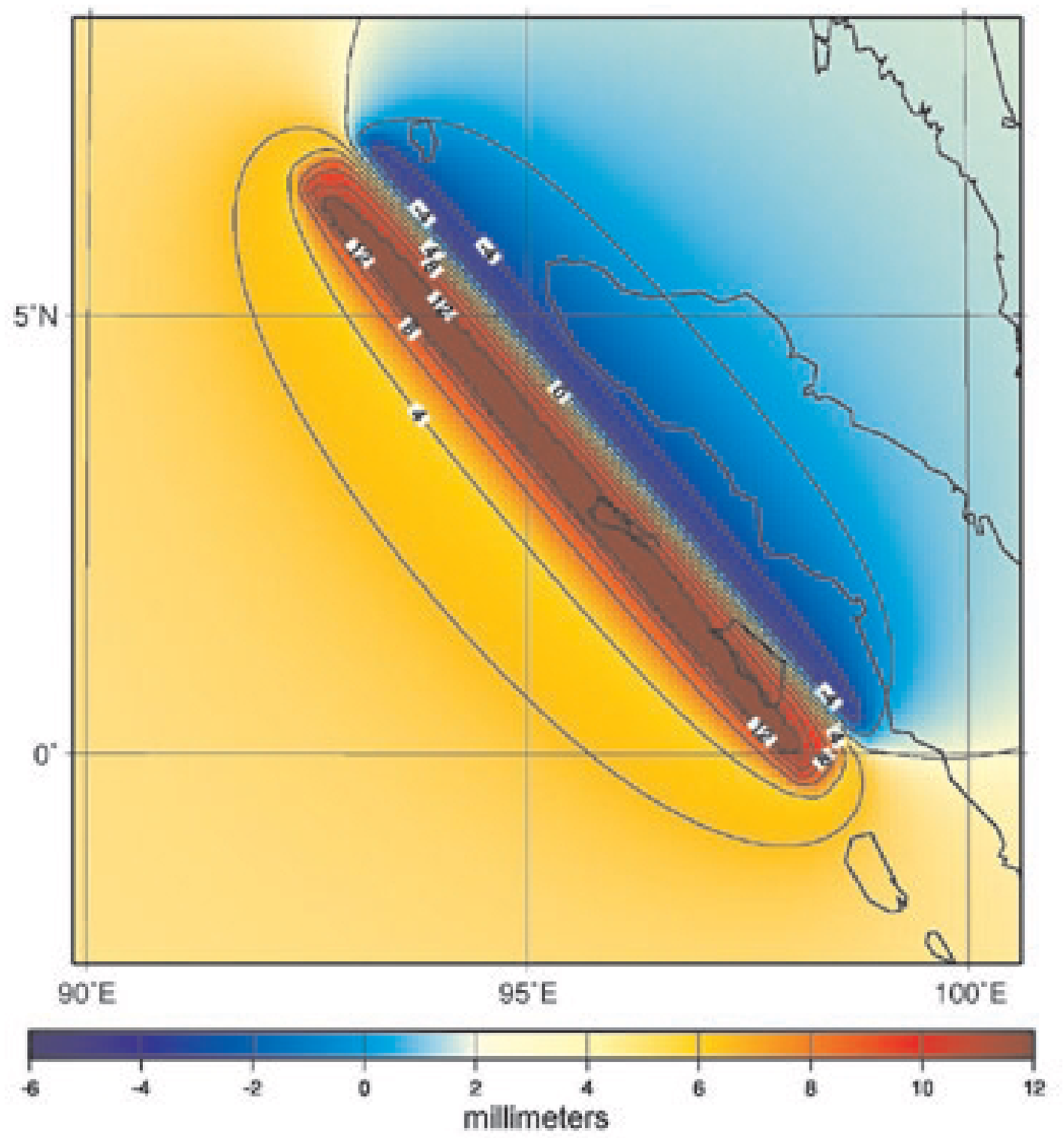,width=60mm,height=60mm,angle=0.0}
   \caption{Geoid deviation estimated from the last Tsunami (on Dec. 26 2004) in Asia, see ref.\cite{GOCE}.  }
   \label{Fig:2Geoid}

   \hspace{3mm}\psfig{figure=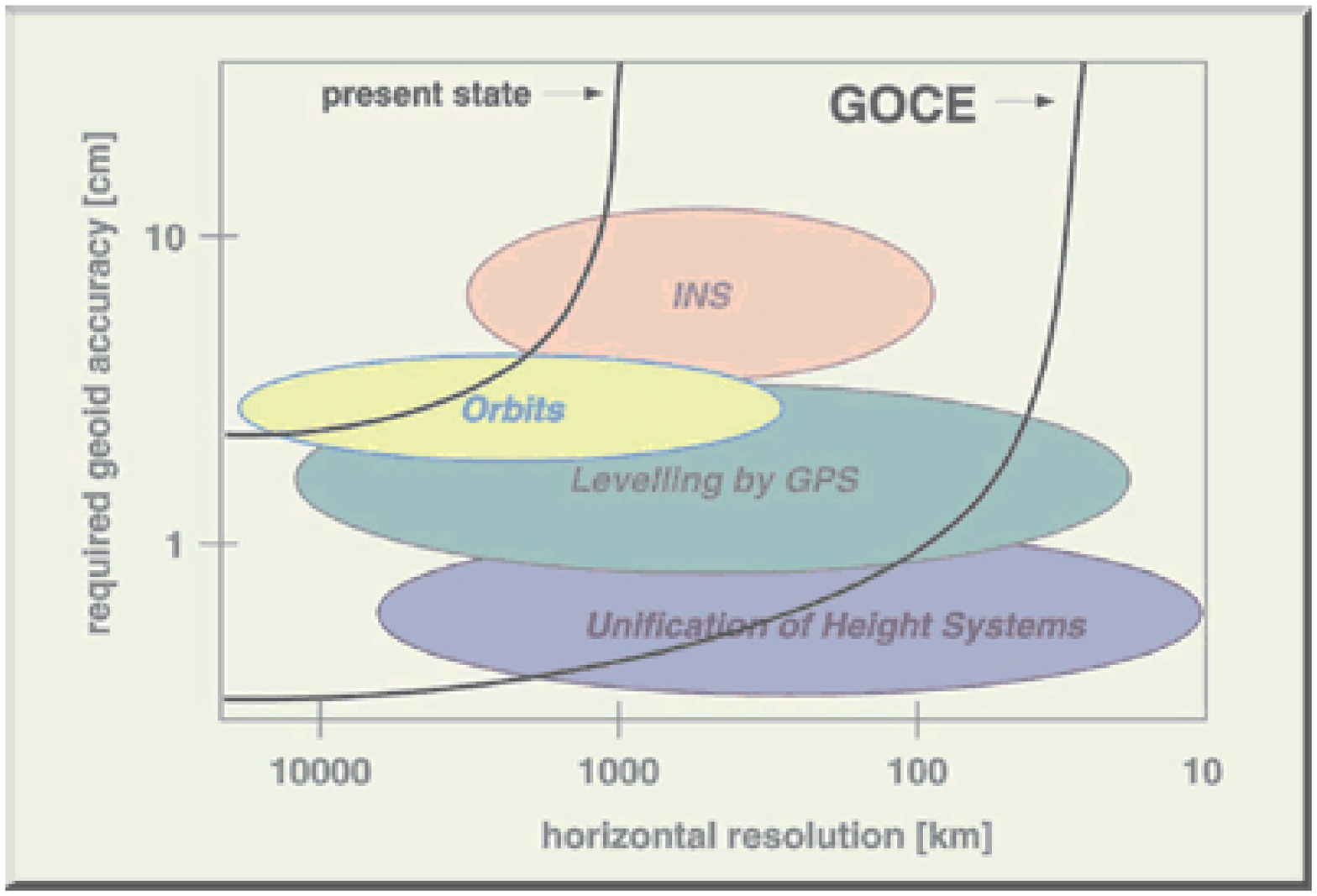,width=60mm,height=60mm,angle=0.0}
   \caption{Geoid deviation accuracy at different horizontal resolution for present
   and future GOCE satellite experiment, see ref. \cite{GOCE}.}
   \label{Fig:Threshold}
   \end{center}
\end{figure}

\section{Conclusions}

  While the recent Tsunami ejected by normal geological ways (P and S waves)
  its huge and lethal message and negligible amount of energy is radiated (as G.W.) at wave zone,
   a very tiny prompt signal took
  place in its near field gravity zone: its prompt tide perturbation
  might be already detectable in principle  by future LISA (if operating at its nominal
  threshold and within $10^4$ km distance and around the Earth ). Unfortunately LIGO and VIRGO antennae
  while being in the same metric  threshold range, are tuned at much higher
  frequencies (tens-thousands Hz). Naturally the  geological waves
  did  over-excite all the terrestrial antennas, but we are dealing
  here  with a immediate gravitational perturbation at light of flight velocity connection.
  Other kind of  Tsunami or Earth-Quake  events  of artificial origin (Appendix C) or by exotic source (Appendix B)
  considered in this paper could  be promptely observed. The  gravity  detection as a prompt event may become the fastest
  alarm system for earliest warn alarm: GOCE satellite \cite{GOCE} (like existing GRACE experiment)
   might become in a near (2006) future
   such a first gravity detection system. Anyway the consequences of such a  huge event is recorded and
   imprinted within our day time length with an increase
    of time a day $ \Delta t_{Tsu} \simeq \mp 2.24 \cdot 10^{-5} \cdot \eta
  s \frac{E_{Tsu}}{10^{27} erg}$. Naturally the worst records are
  imprinted in the tragic human  losses.
\section{Appendix A: Gravitational Wave emission by Tsunami}
  This energy output is apparently a huge and detectable source of gravitational
  waves.  Indeed the approximate G.W. output proportional to the quadrupole third derivative
  square maybe written as :

    $$ \frac{d{E_{GW}}}{dt} \simeq \frac{32}{5} \frac{G}{c^5} \cdot M^2 \cdot r^4 \cdot w^6 = 1.7 \cdot 10^{-10} erg \cdot s^{-1} (\frac{300 s}{t_{Tsun}})^2$$

  where $t_{Tsun} \simeq 3 min.$ is a minimal characteristic
  Tsunami  duration time.
 This gravitational energy power is indeed one of largest
 available on Earth and it is within LIGO (VIRGO-LISA)  frequency range.
 Its output is above $10^{12}$ times larger than the largest mechanical artificial GW source
 as a rotating iron cylinder of $20 m$ lenght and $1$ m diameter at fastest rotation
 angular velocity ($ 28 \frac{rad}{s}$). However even such a power flux at a distance of nearly
  twice the Earth radius,  $10^9$ cm.,  is still many order (more than two decades) of magnitude below the
 LIGO or LISA detection threshold. If this huge energy was been delivered by
 other means (nuclear underground explosion) it was already , as
 discussed below, possible to detect, not by its gravity , but by its prompt neutrino emission.

\section{Appendix B: Anti-meteorite annihilation source of a  Tsunami }

  If an (hypothetical and very un-probable), one-ton, anti-meteorite annihilate on Earth its anti-nucleon
   annihilation will release its ( Richter Magnitude $9$ Tsunami) energy,   mostly in $5$ charged and neutral $400
MeV$ pions (for each nucleon pair annihilation) whose later decay
in flight into $\mu^{\mp}$ and later in
$\nu_e$,$\bar{\nu_e}$,$\nu_{\mu}$,$\bar{\nu_{\mu}}$ will
 inject a prompt (hundreds MeV) neutrino burst  easily detectable by present Super-Kamiokande detector.
 Their signal will be observable ( by more than thousands events)  much better than a  galactic
 supernova and with a quite accurate arrival direction. The energy
 release will produce a wide Tsunami as well a huge atmospheric thunder. Its gravitational wave
 emission power  will be shorter (than previous $3$ minute Tsunami
 time-scale) and the consequent gravitational wave out-put will be  larger by four
 order of magnitude, but still undetectable by present LIGO and VIRGO detectors.
 However the  explosion  itself will occur mostly in air and the explosive destruction will
 propagate anyway at velocity of light to the observable Earth  leading to a prompt un-defensible fire shock-wave.
  Such an energy $10^{27} erg$ output will shine on air  within a few seconds at a luminosity
  (even observed at far horizons within a thousand km distances) comparable at least  to ten thousands
  sun luminosity. The absence of such an events on sun (appearing as a mini-flare)\cite{Solar}, in the past   imply a very
  low probability of such large  anti-meteorite
  existence.\cite{Meteorite}. Very similar result will take place for a
  ton mini-black hole evaporation underground, whose neutrino signal would be
  even harder and very directional.
\section{Appendix C: Underground Nuclear explosion source of a Tsunami ?}

 In case of a $10^{27}$ erg  underground nuclear explosion the large  (MeV) neutrino
   outflow  number $10^{33}$ will be also detectable as a prompt neutrino signal by SK, but at a very
   low energy threshold and without any angular arrival information. On the   contrary the intensity  (hundreds eventS) in a short bomb time (micro-seconds) will be extreme.
   The time correlation between different detectors as SNO,  SK (and others future larger neutrino detectors as UNO) on Earth may offer a very  precise time
   arrival triggering   and distance  triangulation   of the   nuclear explosion location and output calibration. Naturally
     we hope that such an experiment will never take place.
     Anyway the neutrino signal will  be comparable or exceed  any expected galactic supernova
     event \cite{Solar}.
\newcommand\bit{\noindent$\bullet$ \verb}

\end{document}